\documentclass[pra,twocolumn,amsfonts,multicol]{revtex4}

\usepackage{graphicx}
\usepackage{amsmath}
\usepackage{amsfonts}

\newtheorem{lem}{Lemma}
\newtheorem{theo}[lem]{Theorem}

\newcommand{\ket}[1]{|#1\rangle}

\begin{document}
\title{Entanglement swapping between multi-qudit systems}
\author{Jan Bouda$^\dag$ 
 and Vladim\'{\i}r Bu\v{z}ek$^\star$\footnote{
{\bf Permanent address:}
Institute of Physics, Slovak Academy of Sciences,
D\'ubravsk\'a cesta 9, 842 28 Bratislava, Slovakia, and
Faculty of Informatics, Masaryk University, Botanick\'a 68a,
602 00 Brno, Czech Republic }
}
\affiliation{
$^\dag$ Faculty of Informatics, Masaryk University, Botanick\'a 68a, 
602 00 Brno, Czech Republic\\
$^\star$ Department of Physics, University of Queensland, QLD 4072,
Brisbane, Australia
}

\date{20 February 2001}

\begin{abstract}
We generalize the entanglement swapping scheme originally proposed for
two pairs of qubits to an arbitrary number $q$ of systems composed from
an  arbitrary number $m_j$ of qudits. Each of the system is supposed to be
prepared in a maximally entangled state of $m_j$ qudits, while different
systems are not correlated at all.
We show that when a set $\sum_{j=1}^q a_j$ particles 
(from each of the $q$ systems $a_j$ particles are measured) 
is subjected to a generalized Bell-type measurement,
the resulting set of $\sum_{j=1}^q (m_j-a_j)$ particles 
will collapse
into a maximally entangled state. 
\end{abstract}
\maketitle
\section{Introduction}

Recently quantum  entanglement has been
recognized as an important resource for
quantum information processing. In particular,
quantum computation \cite{Gruska99,Nielsen00},
quantum teleportation \cite{Bennett93}, quantum dense coding
\cite{Bennett92},  certain types of quantum key distribution
\cite{Ekert91} and quantum secret sharing protocols \cite{Hillery99}
are rooted in the existence of quantum entanglement.

In spite of all the progress in the understanding of the nature
of quantum entanglement there are still open questions which
have to be answered. In particular, it is not clear yet
how to uniquely quantify the degree of entanglement
\cite{BBPS,Vedral,BDSW,Hill,Horodecki00}, or how to specify
the inseparability conditions for bi-partite multi-level
systems (qudits) \cite{Kraus}. A further problem which
waits for a thorough illumination is the multiparticle entanglement
\cite{Thapliyal99}. There are several aspects of quantum
multiparticle correlations, for instance the investigation
of intrinsic $n$-party entanglement (i.e. generalizations
of the GHZ state \cite{GHZ}). Another aspect of the multiparticle
entanglement is that in contrast to
classical correlation  it cannot freely be shared among many
objects \cite{Coffman,Wootters,Dur00,Koashi00,OConnor00}.

In this paper we want to concentrate our attention on 
entanglement swapping. This  is a method
designed to entangle particles which have never interacted.
The entanglement swapping has been proposed by Zukowski et al. \cite{zuk}
for two pairs of entangled qubits in one of the Bell states. 
Zeilinger et al. \cite{zeil} have generalized the entanglement
swapping to multiparticle systems. Bose et al. \cite{boseswap}
proposed a different version of multiparticle entanglement swapping
and suggested a few interesting ways of using this phenomenon. 
Bose et al. \cite{bose1999} investigated the purification protocol
via  entanglement swapping with non-maximally entangled states.
This approach has been further improved by Shi et al. 
\cite{shi2000}, and Hardy et al. \cite{hardy2000}.   
Delayed choice entanglement swapping has been proposed and analyzed
by Peres \cite{peres2000}.
In \cite{polk,loock} the idea of entanglement swapping has been 
generalized to continuous variables. The use of entanglement
swapping for purification in continuous dimension has  been
proposed by Parker et al. \cite{parker2000}. Entanglement swapping
has been used not only for purification but also for cryptographic
protocols (see, for instance, \cite{cabello2000}).
Finally, we note that
entanglement
swapping has been performed experimentally by Zeilinger et al. \cite{Pan98}.

In this paper we will 
unify all theoretical   approaches to the 
entanglement swapping in one generalized scheme. 
We present 
entanglement swapping for systems consisting of any number of entangled
systems, each composed of an arbitrary number of qudits (i.e. quantum
particles with Hilbert spaces of an arbitrary dimension $D$).
This new unified approach allows us to discuss in detail various scenarios
of multiparticle entanglement. Moreover, our formalism applies to all
possible situations when quantum systems are maximally entangled. 
We do not discuss in this paper entanglement swapping between partially
entangled systems.

In section \ref{maxim} we present a relevant formalism for
a description of kinematics of quantum states in $D$-dimensional
Hilbert spaces.
Section
\ref{pairs} serves as a simple introduction to our swapping
scheme. We show how via a Bell-type measurement entanglement swapping
can be realized. 
This idea is extended in section
\ref{two} for the case of two entangled states, each having
an arbitrary finite number of particles. In section
\ref{general} the most general entanglement scheme  is presented.
We summarize our results in section \ref{concl}.

\section{Entangled qudits}
\label{maxim}

Let the $D$-dimensional Hilbert space be spanned by $D$ orthogonal
normalized vectors $|x_{k}\rangle$, or, equivalently, by $D$ vectors
$|p_{l}\rangle$, $k, l = 0, \ldots, D-1$. These bases are related 
by the discrete Fourier transform
\begin{eqnarray}
\label{2.1}
|x_{k}\rangle &=& \frac{1}{\sqrt{D}} \sum_{l=0}^{D-1} \exp \Bigl(
-i\frac{2\pi}{D}kl \Bigr) |p_{l}\rangle  \, ;
\nonumber \\
|p_{l}\rangle &=& \frac{1}{\sqrt{D}} \sum_{k=0}^{D-1} \exp \Bigl(
i\frac{2\pi}{D}kl \Bigr) |x_{k}\rangle \;.
\end{eqnarray}
Without loss of generality, we assume that these bases are sets 
of eigenvectors of 
two non-commuting operators, the `position' 
$\hat x$ and the `momentum' $\hat p$, such that
\begin{eqnarray}
\label{2.4}
\hat x |x_{k}\rangle = x_k|x_{k}\rangle \;, \quad
\hat p |p_{l}\rangle = p_l|p_{l}\rangle \;,
\end{eqnarray}
where
\begin{eqnarray}
\label{2.5}
x_k= L 
\sqrt{\frac{2\pi}{D}} k ;
\qquad
p_l=\frac{\hbar}{L} \sqrt{\frac{2\pi}{D}} l\;.
\end{eqnarray}
The length, $L$ can, for example, be taken to be
equal to $\sqrt{\hbar/\omega m}$, where $m$ is the mass and $\omega$
the frequency of 
a quantum `harmonic' oscillator within a finite dimensional 
Fock space (in what follows we use units such that
$\hbar=1$).

Next we introduce operators which shift (cyclically permute)
the basis vectors \cite{GTP88}:
\begin{eqnarray}
\label{2.7}
\hat R_{x}(n)|x_{k}\rangle &=& |x_{(k+n){\rm mod}\, D}\rangle\; ; \nonumber \\
\hat R_{p}(m)|p_{l}\rangle &=& |p_{(l+m){\rm mod}\, D}\rangle\;,
\end{eqnarray}
where the sums of indices are taken modulo $D$.
In the $x$-basis these operators can be expressed as
\begin{eqnarray}
\label{2.9}
\langle x_{k}|\hat R_{x}(n)|x_{l} \rangle &=& \delta _{k+n,l}\; ;
\nonumber \\
\langle x_{k}|\hat R_{p}(m)|x_{l} \rangle &=& \delta _{k,l} \exp \Bigl(
i \frac{2 \pi}{D} ml \Bigr) \;.
\end{eqnarray}
Moreover these operators fulfil the Weyl commutation relation 
\cite{Weyl,Sant,Stov}
\begin{eqnarray}
\label{2.10}
\hat R_{x}(n) \hat R_{p}(m) = \exp \Bigl( i \frac{2\pi}{D} mn  \Bigr)
\hat R_{p}(m) \hat R_{x}(n) \; ,
\end{eqnarray}
although they do not commute; they form a representation of an Abelian
group in a ray space. We can displace a state in arbitrary order using
$\hat R_{x}(n) \hat R_{p}(m)$ or $\hat R_{p}(m) \hat R_{x}(n)$, the
resulting state will be the same --- the corresponding kets will differ
only by an unimportant multiplicative factor. We see that  
the operators $\hat R_{x}(n)$ and $\hat R_{p}(m)$ 
displace states in the directions $x$ and $p$, respectively.
The product
$\hat R_{x}(n) \hat R_{p}(m)$ acts as a displacement operator in the
discrete phase space $(k, l)$ \cite{Buz95}. These operators can be 
expressed via the generators of translations (shifts)
\begin{eqnarray}
\label{2.11}
\hat R_{x}(n) &=& 
\exp(-i x_n \hat{p}) \, ;
\nonumber
\\
\hat R_{p}(m) &=& 
\exp(i p_m \hat{x})\;. 
\end{eqnarray}       
We note that the structure of the group associated with the
operators $\hat R_{x}(n)$ and $\hat R_{p}(m)$ is reminiscent of the
group of phase-space translations (i.e., the Heisenberg group)
in quantum mechanics \cite{Fivel95}.

Let us assume a system of two qudits each described by
a vector in a $D$-dimensional Hilbert space ${\cal H}$. 
The tensor product of the two Hilbert spaces can be spanned
by a set of $D^2$ maximally entangled two-qudit states (the analogue
of the Bell basis for spin-$\frac{1}{2}$ particles) 
\cite{Fivel95}
\begin{equation}
\label{2.13}
\begin{split}
|\psi(m,n)\rangle =&\\
=
\frac{1}{\sqrt{D}} &\sum_{k=0}^{D-1} \exp \Bigl(
i\frac{2\pi}{N} mk \Bigr) |x_{k}\rangle|x_{(k-n){\rm mod}\,N}\rangle \,,\!\!\!
\end{split}
\end{equation}
where $m,n=0,\dots,D-1$. These states form an  
orthonormal basis in the space 
${\cal H}\otimes {\cal H}$
\begin{eqnarray}
\label{2.15}
\langle \psi(k,l)|\psi(m,n)\rangle = \delta_{k,m}\delta_{l,n} \;,
\end{eqnarray}
with
\begin{eqnarray}
\label{2.16}
\sum_{m,n=0}^{D-1} |\psi(m,n)\rangle\langle\psi(m,n)| = \hat{I}\otimes
\hat{I}\;.
\end{eqnarray}
In order to prove the above relations we have used the standard relation
$\sum_{n=0}^{D-1}\exp[2\pi i (k-k') n/D] = D \delta_{k,k'}$.

It is interesting to note that the whole set of $D^2$ maximally entangled
states $|\psi(m,n)\rangle$ can be generated from the state
$|\psi(0,0)\rangle$ by the action of {\em local} unitary operations
(shifts) of the form 
\begin{eqnarray}
\label{2.17}
|\psi(m,n)\rangle = 
\hat{R}_p(m)\otimes \hat{R}_x(n)  |\psi(0,0)\rangle \, .
\end{eqnarray}

In what follows we shall simplify our notation. Because we will work 
mostly in the $x$-basis we shall use the notation 
$|x_k\rangle\equiv|k\rangle$. In addition
we will use the notation $x\ominus y$ instead of
$(x-y)\bmod D$. This serves to keep the derivations as synoptical as possible.
Using this notation we can write down the maximally entangled state
of two qudits as 
\begin{eqnarray}
\label{2entangled}
\ket{\psi(l,k)}_{01}=\frac{1}{\sqrt{D}} \sum_{n=0}^{D-1}
e^{\frac{i2\pi ln}{D}}\ket{n}_0\ket{n\ominus k)}_1,
\end{eqnarray}
where parameters $k$ and $l$ can take values between $0$ and $D-1$.

In general, $M$-particle maximally entangled states can be written as
\begin{equation}
\label{multentangled}
\begin{split}
\ket{\Psi}=\ket{\psi(l,k_1,k_2,\dots,k_{M-1})}
=&\\
=\frac{1}{\sqrt{D}}
\sum_{n=0}^{D-1}e^{\frac{i2\pi ln}{D}}\ket{n}_0&\bigotimes_{i=1}^{M-1}
\ket{n\ominus k_i}_i.
\end{split}
\end{equation}
These particles are entangled
in the sense that tracing out any $(M-1)$ particles leaves the reduced density
matrix of the remaining particle  in a maximally mixed state described
by the density operator 
 $\frac{1}{D}I$.

\section{Two entangled pairs}
\label{pairs}

\begin{figure}
\label{fig1}
\begin{center}
\includegraphics{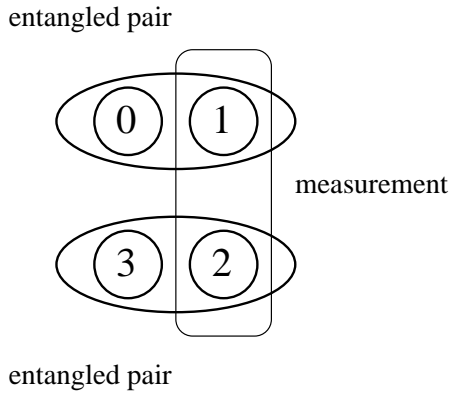}
\caption{
A schematical description of entanglement swapping
between two pairs of qubits (qudits). The qubits 1 and 2 are
measured in the Bell basis. This measurement results in the entanglement
of the qubits 0 and 3 which have never interacted directly.
}
\end{center}
\end{figure}

First of all 
we study a simple example of entanglement swapping
between two qutrits.
Suppose we have two systems each composed of 
two entangled 3-dimensional pairs of particles. The two systems
are not correlated at all and the state vector describing
this composite system can be expressed as
\begin{eqnarray}
\ket{\Psi}&=&\ket{\psi(0,0)}_{01}\otimes\ket{\psi(0,1)}_{23}\\
\nonumber
&=& \frac{1}{\sqrt{3}}\left(\ket{00}_{01}+\ket{11}_{01}+\ket{22}_{01}\right)\\
& &\otimes\frac{1}{\sqrt{3}}\left(\ket{02}_{23}+\ket{10}_{23}+\ket{21}_{23}
\right)
\nonumber
\\
&=&\frac{1}{3}\Big(\ket{00}_{01}\ket{02}_{23}+\ket{00}_{01}\ket{10}_{23}+
\ket{00}_{01}\ket{21}_{23}
\nonumber
\\
&&+\ket{11}_{01}\ket{02}_{23}+\ket{11}_{01}\ket{10}_{23}
+\ket{11}_{01}\ket{21}_{23}
\nonumber
\\
&&+\ket{22}_{01}\ket{02}_{23}+
\ket{22}_{01}\ket{10}_{23}+\ket{22}_{01}\ket{21}_{23}\Big).
\nonumber
\end{eqnarray}
Now assume we
 perform a projective Bell-type 
measurement of particles $1$ and $2$ in the basis 
\eqref{2entangled} with $D=3$. If the measurement yields
$\ket{\psi(r,s)}_{12}$ for some fixed $r$ and $s$, the other two particles
collapse into the state $\ket{\psi(\tilde{l},\tilde{k})}_{03}$.
This result of the measurement conditionally 
`selects' the vectors of the form
\begin{eqnarray}
\ket{n}_0\ket{n}_1\ket{n'}_2\ket{(n'-1)\bmod 3}_3
\end{eqnarray}
for $n=0\dots 2$, such that $n'\equiv n-s\pmod 3$ 
and $\tilde{k}=s+1$. The
amplitude of the vector $\ket{n}_1\ket{(n-s)\bmod 3}_2$
is $e^{i2\pi nr/3}$. It must hold that
\begin{eqnarray}
\label{coefeq}
e^{i2\pi n0/3}e^{i2\pi n'0/3}=e^0=e^{i2\pi nr/3}
e^{i2\pi n\tilde{l}/3}.
\end{eqnarray}
Since $e^{ui2\pi}=e^{u'i2\pi}\,\forall u,u'\in \mathbb Z$ the equation
\eqref{coefeq} holds for $\tilde{l}=(-r)\bmod 3$.
The previous derivations yield that the state of the
particles $0$ and $3$ collapses into 
the maximally entangled state
$\ket{\psi((0-r)\bmod 3,(s+1)\bmod 3)}_{03}$
of two qutrits.

\subsection*{Measuring a general state}

Let us consider now a slightly more complex situation. We have a system
of two entangled pairs in the general state
$\ket{\psi(l,k)}_{01}\otimes\ket{\psi(l',k')}_{23}$. When we perform
the measurement according to the basis \eqref{2entangled} with $D=3$ we obtain
the vector $\ket{\psi(r,s)}_{12}$.
The resulting state of particles $0$ and $3$ is again  denoted as 
$\ket{\psi(\tilde{l},\tilde{k})}_{03}$. In this case we are looking for the
vectors of the form
\begin{eqnarray}
\label{3genvec}
\ket{n}_0\ket{n\ominus k}_1\ket{n'}_2\ket{n'\ominus k'}_3
\end{eqnarray}
such that $n'\equiv n-k-s\pmod 3$, which yields $\tilde{k}=(k+s+k')\bmod 3$.
The coefficient of the
vector $\ket{n\ominus k}_1\ket{n\ominus k\ominus s}_2$
is $e^{i2\pi (n-k)r/3}$. It must hold as before (see equation \eqref{coefeq})
 that
\begin{equation}
e^{i2\pi nl/3}e^{i2\pi (n-k-s)l'/3}=e^{i2\pi (n-k)r/3}
e^{i2\pi n\tilde{l}/3}e^{i2\pi x /3}
\end{equation}
for $n=0,1,2$, where $e^{i2\pi x/3}$ will be part of the phase shift
of the vector $\ket{\psi(\tilde{l},\tilde{k})}_{03}$.
This implies the congruence
\begin{equation}
\label{lfind}
n(l+l'-r-\tilde{l})\equiv-kr+kl'+sl'+x\pmod 3.
\end{equation}
The case $n=0$ gives $-kr+kl'+sl'+x\equiv0\pmod 3$, so the $x$ must be chosen
such that 
 this congruence is satisfied. For $n=1$ this leads to a 
relation 
\begin{eqnarray}
\label{lfound}
\tilde{l}=(l+l'-r)\bmod 3.
\end{eqnarray}
The extension to an arbitrary finite-dimensional systems is straightforward.
It suffices to replace all `$\!\!\mod 3$' by 
`$\!\!\mod D$' and $n$ varies from
$0$ to $D-1$. In equation \eqref{3genvec} the generalization to
$D$-dimensional system gives us $n'\equiv n-k-s\pmod D$. Since 
$n$ varies from $0$ to $D-1$, we have D vectors of the form \eqref{3genvec}.
Therefore their linear combination with appropriate coefficients
gives $\ket{\psi(\tilde{l},\tilde{k})}_{03}$ and not only a linear combination
of less than $D$ distinct vectors of the form $e^{i2\pi \tilde{l}n/D}\ket{n}_0
\ket{n\ominus \tilde{k}}_3$.
We can now summarize our results as follows.
\begin{theo}
Suppose that $\ket{\Psi}=\ket{\psi(l,k)}_{01}\otimes\ket{\psi(l',k')}_{23}$ is
the tensor product of two  maximally entangled pairs 
of qudits. Let assume that the
particles $1$ and $2$ are measured via the Bell-type measurement 
in the basis \eqref{2entangled}.
If the measurement yields the result $\ket{\psi(r,s)}_{12}$, then 
the two particles $0$ and $3$ collapse into the state
\begin{eqnarray}
\ket{\psi((l+l'-r)\bmod D,(k+k'+s)\bmod D)}_{03}.
\end{eqnarray}
This is a maximally entangled state of qudits $0$ and $3$, which have
never interacted before.
\end{theo}

\section{Entangling two multiparticle systems}
\label{two}

\subsection*{Measurement of  two particles}
\begin{figure}
\label{fig2}
\begin{center}
\includegraphics{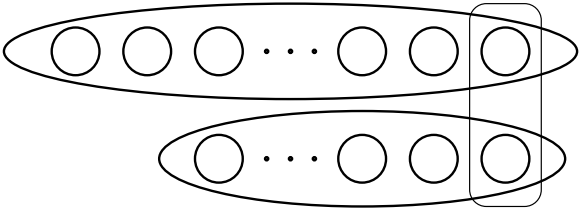}
\caption{
A schematical description of entanglement swapping
between two sets of entangled qudits. A single particle
from each set is measured.
 This measurement results in entanglement
between the rest of the particles from both of the
systems.
}
\end{center}
\end{figure}

Suppose we have two uncorrelated systems of qudits. The first
system with $m_1+1$ qudits is in a maximally entangled state
$\ket{\psi(l,k_1,\dots,k_{m_1})}$, while the second
system with $m_2+1$ qubits is in the state
$\ket{\psi(l',k'_1,\dots,k'_{m_2})}$. The state vector of
the composite system then reads
$\ket{\psi(l,k_1,\dots,k_{m_1})}\otimes\ket{\psi(l',k'_1,\dots,k'_{m_2})}$.
Now we can choose two arbitrary particles  
(one from each of the two systems) 
to be measured using the
Bell-type projective measurement. 
Due to the cyclic symmetry  we can assume
that the `last' particle of each of the two systems is measured.
Suppose that in a measurement 
we obtain a  state
$\ket{\psi(r,s)}$.
Therefore we are looking for vectors of the form
\begin{eqnarray}
\ket{n}\ket{n\ominus k_1}\dots\ket{n\ominus k_{m_1}}
\ket{n'}\dots \ket{n'\ominus k'_{m_2}}
\end{eqnarray}
such that $n'\equiv n-k_{m_1}-s+k'_{m_2}\pmod D$. To simplify the following
derivations we  put $k_0=k'_0=0$. Let
\begin{eqnarray}
\Delta\tilde{k}=k_{m_1}+s-k'_{m_2}.
\end{eqnarray}
Now we should determine the $\tilde{l}$ and therefore 
\begin{equation}
\begin{split}
e^{i2\pi nl/D}e^{i2\pi (n-k_{m_1}-s+k'_{m_2})l'/D}=&\\
=e^{i2\pi (n-k_{m_1})r/D}&e^{i2\pi n\tilde{l}/D}
e^{i2\pi x/D}.
\end{split}
\end{equation}
It follows that
\begin{equation}
\begin{split}
n(l+l'-r-\tilde{l})\equiv&\\
\equiv -k_{m_1}r+k_{m_1}&l'+sl'-k'_{m_2}l' 
+ x\pmod D.
\end{split}
\end{equation}
As before (see equations \eqref{lfind} and \eqref{lfound}) 
for $n=1$ we have
\begin{eqnarray}
\tilde{l}=(l+l'-r)\bmod D.
\end{eqnarray}
Once we have determined $\tilde{l}$, we can choose suitable $x$ to satisfy
the case $n=0$. This means the following congruence is equal to zero:
\begin{equation}
-k_{m_1}r+k_{m_1}l'+sl'-k'_{m_2}l'+x\equiv0\pmod D.
\end{equation}

The resulting state is
$
\ket{\psi(\tilde{l},\tilde{k}_1,\dots,\tilde{k}_{m_1+m_2-1})},
$
where
\begin{align*}
\tilde{k}_i&=k_i \, & i<m_1 \, \\
\tilde{k}_i&=k'_{i-m_1} + \Delta\tilde{k}\,  &
m_1\le \ i\, .
\end{align*}
In this section we have  presented a technique
which allows us to produce an entangled
state with any number of particles.

\subsection*{Measuring more than two particles}
\begin{figure}
\label{fig3}
\begin{center}
\includegraphics{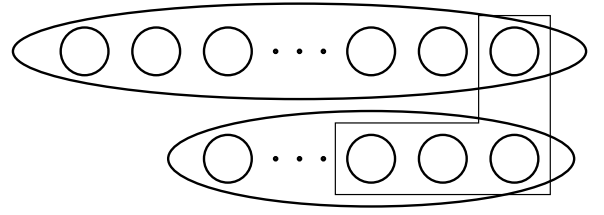}
\caption{
The same as in figure 2 except an arbitrary number 
of particles from each set is measured.
}
\end{center}
\end{figure}
Suppose that we are measuring the
last $a_1$ particles of the first system  and the last $a_2$
particles of the second system. We again assume the Bell-type
measurement in  the basis
$\ket{\psi(r,s_1,\dots,s_{a_1+a_2-1})}$ describing 
 maximally entangled states of the $a_1+a_2$ qudits.

Analogically as in the previous examples we are looking for vectors of 
the form
\begin{equation}
\begin{split}
\ket{n}\dots\ket{n\ominus k_{m_1-a_1+1}}\dots\ket{n\ominus k_{m_1}}\otimes&
\\
\otimes\ket{n'}\dots\ket{n'\ominus k'_{m_2-a_2+1}}\dots&\ket{n'\ominus k_{m_2}}
\end{split}
\end{equation}
such that $n'\equiv n-k_{m_1-a_1+1}-s_{a_1}+k_{m_2-a_2+1}$
for a given result of the measurement
$\ket{\Psi}=\ket{\psi(r,s_1,\dots,s_{a_1+a_2-1})}$.
Let $\Delta\tilde{k}=k_{m_1-a_1+1}+s_{a_1}-k_{m_2-a_2+1}$ and $k_0'=0$.
Now let us determine $\tilde{l}$. It holds that
\begin{equation}
\begin{split}
e^{i2\pi (nl-x)/D}e^{i2\pi (n-k_{m_1-a_1+1}-s_{a_1}+k'_{m_2-a_2+1})l'/D}=\\
=e^{i2\pi (n-k_{m_1-a_1+1})r/D}e^{i2\pi n\tilde{l}/D}.
\end{split}
\end{equation}
This leads again to the relation 
\begin{eqnarray}
\tilde{l}=(l+l'-r)\bmod D,
\end{eqnarray}
so the state of the unmeasured particles is
$
\ket{\psi(\tilde{l},\tilde{k}_1,\dots,\tilde{k}_{m_1+m_2+1-a_1-a_2})},
$
where
\begin{eqnarray}
\tilde{k}_i&=k_i\, \qquad &i<m_1-a_1+1\,
\nonumber
\\
\tilde{k}_i&=k'_{i-m_1+a_1-1}+\Delta\tilde{k}\, \qquad &m_1-a_1+1\le\ i.
\nonumber
\end{eqnarray}
\begin{theo}
Suppose that we have two entangled systems 
with $m_1+1$ and $m_2+1$ particles, respectively, 
initially prepared in the state
\begin{eqnarray}
\ket{\Psi}=\ket{\psi(l,k_1,\dots,k_{m_1}}\otimes
\ket{\psi(l',k'_1,\dots,k'_{m_2}}
\end{eqnarray}
and suppose 
that we subject the last $a_1$ particles from the first system and the last 
$a_2$ particles from the second system to a joint Bell-type 
measurement in the basis formed by vectors
$
\ket{\psi(r,s_1,\dots,s_{a_1+a_2-1})}.
$
Then the vector describing the state of the remaining
$m_1+m_2+2-a_1-a_2$  particles after the measurement is
\begin{eqnarray}
\ket{\psi(\tilde{l},\tilde{k}_1,\dots,\tilde{k}_{m_1+m_2+1-a_1-a_2})}
\end{eqnarray}
where
\begin{eqnarray}
\tilde{k}_i=k_i& \, \qquad &i<m_1-a_1+1\, 
\nonumber
\\
\tilde{k}_i=k'_{i-m_1+a_1-1}+\Delta\tilde{k}&\, \qquad &m_1-a_1+1\le i
\nonumber
\\
\tilde{l}=(l+l'-r)\bmod D.&&
\nonumber
\end{eqnarray}
This means that the remaining particles end up in a maximally entangled state.
\end{theo}

\section{Many multiparticle entangled states}
\begin{figure}
\label{fig4}
\begin{center}
\includegraphics{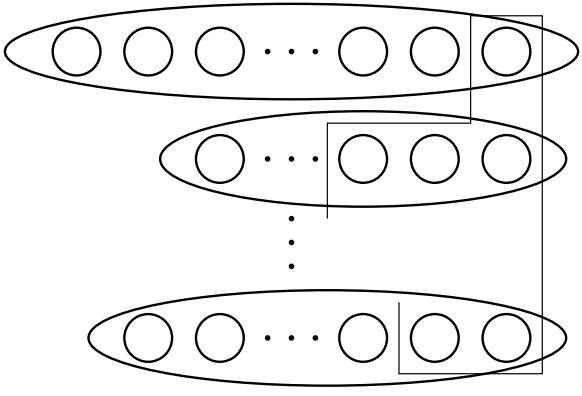}
\caption{
The same as in figure 3 except that many initially uncorrelated 
multi-qudit systems 
are considered.
}
\end{center}
\end{figure}
\label{general}
In what follows  we describe
the most general situation for entanglement swapping:
Suppose we have $q$ systems. The $j$th system is composed
of $m_j+1$ ($j=1,\dots,q$) particles which are in a maximally
entangled state $\ket{\psi(l^j,k_1^j,\dots,k_{m_j}^j)}$.
The different systems are totally factorized, so the 
state vector of the composite system reads
\begin{eqnarray}
\label{genstate}
\ket{\Psi}=\bigotimes_{j=1}^q \ket{\psi(l^j,k_1^j,\dots,k_{m_j}^j)}.
\end{eqnarray}
(We note that  superscripts do not denote the power, 
but they serve  as indices.)
Further we assume a multiparticle Bell-type  measurement. Specifically,
we consider 
$a_j$ particles from $j$th state,
$\forall j\in{1\dots q}$, to be measured simultaneously 
 in the basis
\begin{eqnarray}
\label{basis}
\ket{\psi(r,s_1^1,\dots,s_{a_1}^1,s_1^2,\dots,s_{a_q-1}^q)}.
\end{eqnarray}
The total number of measured particles is $\sum_{j=1}^q a_j$.
After the measurement these particles collapse into one of the
vectors \eqref{basis}.
Therefore we look for 
the vectors 
\begin{eqnarray}
\ket{n^1}\ket{n^1\ominus k_1^1}\dots\ket{n^1\ominus k^1_{m_1-a_1+1}}\dots
\ket{n^1\ominus k^1_{m_1}}
\nonumber
\\
\otimes\dots\otimes\ket{n^q}\dots
\ket{n^q\ominus k^q_{m_q-a_q+1}}\dots\ket{n^q\ominus k^q_{m_q}}
\end{eqnarray}
such that
\begin{eqnarray}
n^2&\equiv n^1-k^1_{m_1-a_1+1}-s^1_{a_1}+k^2_{m_2-a_2+1}\pmod D
\nonumber
\\
n^3&\equiv n^1-k^1_{m_1-a_1+1}-s^2_{a_2}+k^3_{m_3-a_3+1}\pmod D
\\
&\vdots& \, ,
\nonumber
\end{eqnarray}
which in general can be expressed as
\begin{equation}
n^i\equiv n^1-k_{m_1-a_1+1}-s^{i-1}_{a_{i-1}}+k^i_{m_i-a_i+1}\pmod D
\end{equation}
 for $\forall i=2\dots q$.
It remains to determine $\tilde{l}$.
As before we have 
\begin{equation}
\prod_{j=1}^q e^{i2\pi n^jl^j/D}=e^{i2\pi(n^1-k_{m_1-a_1+1})r/D}
e^{i2\pi n^1\tilde{l}/D} e^{i2\pi x/D},
\end{equation}
which yields
\begin{equation}
\begin{split}
\label{40}
n^1\left(\left(\sum_{j=1}^q l^j\right)-r-\tilde{l}\right)\equiv
-k_{m_1-a_1+1}r+x
\\
+
\sum_{j=2}^q l^j\left(k^1_{m_1-a_1+1}+s^{j-1}_{a_{j-1}}-k^j_{m_j-a_j+1}\right)
\pmod D.
\end{split}
\end{equation}
The right-hand side of the congruence \eqref{40} is equal to
$0\pmod D$ which affects only the global phase.
Therefore we can write 
\begin{eqnarray}
\tilde{l}=\left(\left(\sum_{j=1}^q l^j\right)-r\right)\mod D.
\end{eqnarray}
Consequently a set of $\sum_j(m_j-a_j+1)$ unmeasured particles
becomes entangled due to the Bell-type measurement performed on
the $\sum_j a_j$ particles. 
The state of the unmeasured particles is 
\begin{eqnarray}
\ket{\psi(\tilde{l},\tilde{k}^1_1,\dots,\tilde{k}^1_{m_1-a_1+1},\tilde{k}^2_1,
\dots,\tilde{k}^q_{m_q-a_q})}.
\end{eqnarray}
Together there are $\left(\sum_{j=1}^q m_j-a_j+1\right)-1$
$\tilde{k}$ and they must satisfy the condition
\begin{eqnarray}
\tilde{k}_i^j&=&k_i^j+n^j-n^1\, \qquad i\le m_j-a_j
\nonumber
\\
\tilde{k}_{m_j-a_j+1}^j&=&n^{j+1}-n^1.
\end{eqnarray}
\begin{theo}
Suppose we have $q$ entangled systems 
each composed of $m_j+1$ particles ($j=1,\dots,q$). Let the
whole system is initially in the state
\eqref{genstate}. Let us subject the last $a_j$ particles from $j$-th
($\forall j=1\dots q$) system to the 
Bell-type measurement in the basis formed by vectors
\eqref{basis}. Given the result of the measurement 
\eqref{basis} 
the $\sum_j(m_j-a_j+1)$ 
unmeasured particles collapse into the
maximally entangled  state
\begin{eqnarray}
\ket{\psi(\tilde{l},\tilde{k}^1_1,\dots,\tilde{k}^1_{m_1-a_1+1},\tilde{k}^2_1,
\dots,\tilde{k}^q_{m_q-a_q})},
\end{eqnarray}
where
\begin{eqnarray}
\tilde{l}=\left(\left(\sum_{j=1}^q l^j\right)-r\right)\mod D
\end{eqnarray}
and
\begin{align*}
\tilde{k}_i^j&=k_i^j+n^j-n^1\, ;& &i\le m_j-a_j;\\
\tilde{k}_{m_j-a_j+1}^j&=n^{j+1}-n^1.
\end{align*}
\end{theo}

\section{Conclusion}
\label{concl}
In this paper we have presented a general formalism describing
entanglement swapping between multi-qudit systems. We have
shown that by performing Bell-type measurements
one can 
create entangled states (with
an arbitrary number of particles) from particles which have never interacted
before. 

Even though our formalism has been developed for finite-dimensional
Hilbert space, it can be generalized  for continuous
variables, i.e. $D\rightarrow\infty$. In this case qudits are
replaced by harmonic oscillators (e.g. quantized modes of an
electromagnetic field). Formally, in the limit
$D\rightarrow\infty$ we can substitute a two-qudit maximally
entangled state by a two-mode correlated state, i.e.
\begin{eqnarray}
\label{sq}
\frac{1}{\sqrt{D}}\sum_n e^{ip_l x_n}\ket{x_n}
\ket{x_n-x_k} \rightarrow |\psi(x,p)\rangle
\end{eqnarray}
where
\begin{eqnarray}
|\psi(x,p)\rangle\equiv\frac{1}{\sqrt{2\pi}}
\int d \tilde{x}  e^{i p\tilde{x}}|\tilde{x}\rangle_0
|\tilde{x}-x\rangle_1.
\end{eqnarray}
Analogously, a multi-mode entangled state in the continuous
limit can be expressed as
\begin{eqnarray}
\label{sqm}
\frac{1}{\sqrt{2\pi}}\int d\tilde{x}
\ e^{ip\tilde{x}}\ket{\tilde{x}}_0\bigotimes_{j=1}^{M-1}
\ket{\tilde{x}-x_j}_j.
\end{eqnarray}
Once these states are defined one can formally perform the
same manipulations as in the case of qudits, i.e.
generalized Bell measurements, etc. Nevertheless, we remind
ourselves that the maximally correlated states \eqref{sq}
as well as \eqref{sqm} require infinite energy for their
creation. For this reason it is desirable to consider 
two-mode (and multi-mode) squeezed states which in the limit
of infinite squeezing are equal to \eqref{sq} and 
\eqref{sqm}, respectively. It is convenient to describe
these two mode state in term of their Wigner functions.
In particular, the Wigner function corresponding to a 
regularized version of the state $\ket{\psi(0,0)}$ 
is \cite{Braunstein98}
\begin{equation}
\begin{split}
\label{2.33}
W(x_1,p_1;x_2,p_2) 
&= \exp\left\{
-\frac{e^{2\xi}}{2}\left[(x_1-x_2)^2 + (p_1+p_2)^2\right]
\right\}
\\
&\times 
\exp\left\{
-\frac{e^{-2\xi}}{2}\left[(x_1+x_2)^2 + (p_1-p_2)^2\right]
\right\} \;.
\end{split}
\end{equation}
This is a Wigner function describing a two-mode squeezed vacuum. If
we trace over one of the modes, i.e., if we perform an integration over
the parameters $x_2$ and $p_2$ we obtain from (\ref{2.33}) a Wigner
function of a thermal field                                            
where $\bar{n}=\sinh^2\xi$ is the mean excitation number in the two-mode
squeezed vacuum under consideration. We note that the thermal state
 is a maximally mixed state (i.e. the state with the
highest value of the von Neumann entropy) for a given mean excitation
number. This means that the pure state (\ref{2.33}) is the most
entangled state for a given mean excitation number.
From this it follows that to create a truly maximally entangled state,
i.e. the state (\ref{2.33}) in the limit $\xi\rightarrow\infty$, an
infinite number of quanta is needed and so infinite energy.   

From \eqref{2.33} one can easily find the Wigner functions
of other states $\ket{\psi(x,p)}$. We remind ourselves that
Wigner functions are invariant under canonical transformations
\eqref{2.11}. Taking into account that 
 states $\ket{\psi(x,p)}$
can be obtained from $\ket{\psi(0,0)}$ by a canonical transformation
(see \eqref{2.17})
\begin{eqnarray}
\ket{\psi(x,p)} = \hat{R}_p(p)\otimes \hat{R}_x(x)\ket{\psi(0,0)}
\end{eqnarray}
its Wigner function can be obtained via a 
simple substitution of variables from
the Wigner function \eqref{2.33}. 
The generalized Bell measurement in this representation corresponds
to a POVM measurement of the Artur-Kelly type \cite{Buz95}.
 This formalism
in the infinite squeezing then leads to a perfect entanglement
swapping between harmonic oscillators.

{\bf Acknowledgements}\newline
We thank 
professor Jozef Gruska for  stimulating discussions.
This work was supported by the IST project EQUIP under the contract
IST-1999-11053 and by GA\v CR grant 201/98/0369.
 VB acknowledges support from the University of Queensland
Traveling Scholarship.

\bigskip


\end{document}